\documentclass{aastex631}

\usepackage{amsmath}
\usepackage{graphicx}
\usepackage{natbib}
\begin{document}

\title{Reanalysis of the X-ray burst associated FRB 200428 with \textit{Insight}-HXMT observations}

\author{M.Y. Ge}
\affiliation{Key Laboratory of Particle Astrophysics, Institute of High Energy Physics, Chinese Academy of Sciences, Beijing 100049, China}

\author{C.Z. Liu\textsuperscript{*}}
\affiliation{Key Laboratory of Particle Astrophysics, Institute of High Energy Physics, Chinese Academy of Sciences, Beijing 100049, China}

\author{S.N. Zhang}
\affiliation{Key Laboratory of Particle Astrophysics, Institute of High Energy Physics, Chinese Academy of Sciences, Beijing 100049, China}
\affiliation{University of Chinese Academy of Sciences, Chinese Academy of Sciences, Beijing 100049, China}

\author{F.J. Lu}
\affiliation{Key Laboratory of Particle Astrophysics, Institute of High Energy Physics, Chinese Academy of Sciences, Beijing 100049, China}
\affiliation{Key Laboratory of Stellar and Interstellar Physics and School of Physics and Optoelectronics,Xiangtan University, Xiangtan 411105, China}

\author{Z. Zhang}
\affiliation{Key Laboratory of Particle Astrophysics, Institute of High Energy Physics, Chinese Academy of Sciences, Beijing 100049, China}

\author{Z. Chang}
\affiliation{Key Laboratory of Particle Astrophysics, Institute of High Energy Physics, Chinese Academy of Sciences, Beijing 100049, China}

\author{Y.L. Tuo}
\affiliation{Key Laboratory of Particle Astrophysics, Institute of High Energy Physics, Chinese Academy of Sciences, Beijing 100049, China}

\author{X.B. Li}
\affiliation{Key Laboratory of Particle Astrophysics, Institute of High Energy Physics, Chinese Academy of Sciences, Beijing 100049, China}

\author{C.K. Li}
\affiliation{Key Laboratory of Particle Astrophysics, Institute of High Energy Physics, Chinese Academy of Sciences, Beijing 100049, China}

\author{S.L. Xiong}
\affiliation{Key Laboratory of Particle Astrophysics, Institute of High Energy Physics, Chinese Academy of Sciences, Beijing 100049, China}

\author{C. Cai}
\affiliation{Key Laboratory of Particle Astrophysics, Institute of High Energy Physics, Chinese Academy of Sciences, Beijing 100049, China}

\author{X.F. Li}
\affiliation{Key Laboratory of Particle Astrophysics, Institute of High Energy Physics, Chinese Academy of Sciences, Beijing 100049, China}

\author{R. Zhang}
\affiliation{Theoretical Physics Division, Institute of High Energy Physics, Chinese Academy of Sciences, \\
19B Yuquan Road, Beijing 100049, People’s Republic of China}

\author{Z.G. Dai}
\affiliation{CAS Key Laboratory for Research in Galaxies and Cosmology, Department of Astronomy, University of Science and Technology of China, Hefei 230026, China}

\author{J.L. Qu}
\affiliation{Key Laboratory of Particle Astrophysics, Institute of High Energy Physics, Chinese Academy of Sciences, Beijing 100049, China}

\author{L.M. Song}
\affiliation{Key Laboratory of Particle Astrophysics, Institute of High Energy Physics, Chinese Academy of Sciences, Beijing 100049, China}

\author{S. Zhang}
\affiliation{Key Laboratory of Particle Astrophysics, Institute of High Energy Physics, Chinese Academy of Sciences, Beijing 100049, China}

\author{L.J. Wang}
\affiliation{Key Laboratory of Particle Astrophysics, Institute of High Energy Physics, Chinese Academy of Sciences, Beijing 100049, China}

\begin{abstract}
A double-peak X-ray burst from the Galactic magnetar SGR J1935+2154 was discovered as associated with the two radio pulses of FRB 200428 separated by $28.97\pm0.02$\,ms. Precise measurements of the timing and spectral properties of the X-ray bursts are helpful for understanding the physical origin of fast radio bursts (FRBs). In this paper, we have reconstructed some information about the hard X-ray events, which were lost because the High Energy X-ray Telescope (HE) onboard the \textit{Insight}-HXMT mission was saturated by this extremely bright burst, and used the information to improve the temporal and spectral analyses of the X-ray burst. The arrival times of the two X-ray peaks by fitting the new \textit{Insight}-HXMT/HE lightcurve with multi-Gaussian profiles are $2.77\pm0.45$\,ms and $34.30\pm0.56$\,ms after the first peak of FRB 200428, respectively, while these two parameters are $2.57\pm0.52$\,ms and $32.5\pm1.4$\,ms if the fitting profile is a fast rise and exponential decay function. The spectrum of the two X-ray peaks could be described by a cutoff power-law with cutoff energy $\sim{60}$\,keV and photon index $\sim{1.4}$, the latter is softer than that of the underlying bright and broader X-ray burst when the two X-ray peaks appeared.

\end{abstract}


\section{Introduction}

Fast radio bursts (FRBs) are the brightest flashes in the radio sky \citep{2019A&ARv..27....4P,2019ARA&A..57..417C}. However, their physical mechanism has remained mysterious so far \citep{2020Natur.587...45Z,2021SCPMA..6449501X}, although the discovery of FRB 200428 establishes magnetars as an origin of some FRBs \citep{2020Natur.587...59B,2020Natur.587...54C}. At the same time, its X-ray counterpart, a bright X-ray burst (XRB), was detected by high energy instruments such as {\sl Insight}-HXMT, INTEGRAL, Konus-{\sl Wind}, and AGILE \citep{2021NatAs...5..378L,2020ApJ...898L..29M,2021NatAs...5..372R,2021NatAs...5..401T}. {\sl Insight}-HXMT discovered the double X-ray peaks corresponding to the double radio peaks \citep{2021NatAs...5..378L}, and both {\sl Insight}-HXMT and INTEGRAL localized the X-ray burst as coming from SGR~J1935+2154 \citep{2021NatAs...5..378L,2020ApJ...898L..29M}. It is the first time that a counterpart of an FRB was detected and localized at other wavelengths, which allowed the identification of the origin of an FRB
\citep{2022ATel15667,2022ATel15690,2022ATel15672,2022ATel15698}.

Since October 2022, SGR J1935+2154 went into a new active episode \citep{2022ATel15667,2022ATel15690,2022ATel15672,2022ATel15698}. After the burst forest, at least two FRB signals were captured by CHIME \citep{2022ATel15681....1D}, GBT \citep{2022ATel15697....1M} and Yunnan 40\,m radio telescope (ATEL 15707) and their corresponding X-ray burst are detected by GECAM, HEBS(GECAM-C), Konus-Wind and {\sl Insight}-HXMT \citep{2022ATel15682,2022ATel15686....1F,2022ATel15708}. Compared to these results, more constraints on the FRB mechanism could be supplied by the timing properties of FRB 200428 and XRB, such as the time delay and the quasi-period oscillation in XRB \citep{2021NatAs...5..378L,2021NatAs...5..372R,2020ApJ...898L..29M,2022ApJ...931...56L}. In the previous study, marginal time delay in the XRB lightcurves has been detected by {\sl Insight}-HXMT, especially for the second peak \citep{2021NatAs...5..378L}. \cite{2020ApJ...898L..29M} also report that there is a hard X-ray time delay of $6.5\pm1.0$\,ms for the second peak. Until now, no more measurements have been reported for the timing properties between FRB 200428 and the XRB. We have recovered some of the hard X-ray events of this XRB, which were lost because the High Energy X-ray Telescope (HE) onboard the \textit{Insight}-HXMT mission was saturated by this extremely bright burst. Here, we utilize the recovered data of {\sl Insight}-HXMT/HE together with {\sl Insight}-HXMT/ME, INTEGRAL and Konus-{\sl Wind} to perform a detailed timing analysis for the double-peak structure of the XRB by fitting its lightcurve.

\section{Observations and Data processing}

{\it Insight}-HXMT is China's first X-ray astronomical satellite, which carries three main payloads: the High Energy X-ray telescope (HE, 20-250\,keV, 5100\,cm$^{2}$), the Medium Energy X-ray telescope (ME, 5-30\,keV, 952\,cm$^{2}$), and the Low Energy X-ray telescope (LE, 1-15\,keV, 384\,cm$^{2}$) \citep{1993Ap&SS.206...91L,1993Ap&SS.205..381L,2014SPIE.9144E..21Z,2020SCPMA..6349502Z,2020SCPMA..6349503L,2020SCPMA..6349504C,2020SCPMA..6349505C}. {\it Insight}-HXMT was in pointing observation mode and acquired high quality data when FRB 200428 occurred. The data supply a unique opportunity to study the timing properties of the FRB event in X-ray energy band. The {\sl Insight}-HXMT observation ID including the XRB corresponding to FRB 200428 is P0314003001. The data process of XRB is similar to the procedure described as in \cite{2021NatAs...5..378L}, which only used the recovered lightcurve of one third of the detectors of {\it Insight}-HXMT/HE (HE for clarity, hereafter). In this work, all the detectors of HE are used to perform timing analysis. The results of {\it Insight}-HXMT/ME (ME for clarity, hereafter) are obtained from \cite{2021NatAs...5..378L} directly. However, the results of {\it Insight}-HXMT/LE (LE for clarity, hereafter) are not utilized here as the first peak is not detected in LE lightcurve \citep{2021NatAs...5..378L}.

\begin{figure}[!htb]
  \centerline{
      \includegraphics[width=0.6\columnwidth]{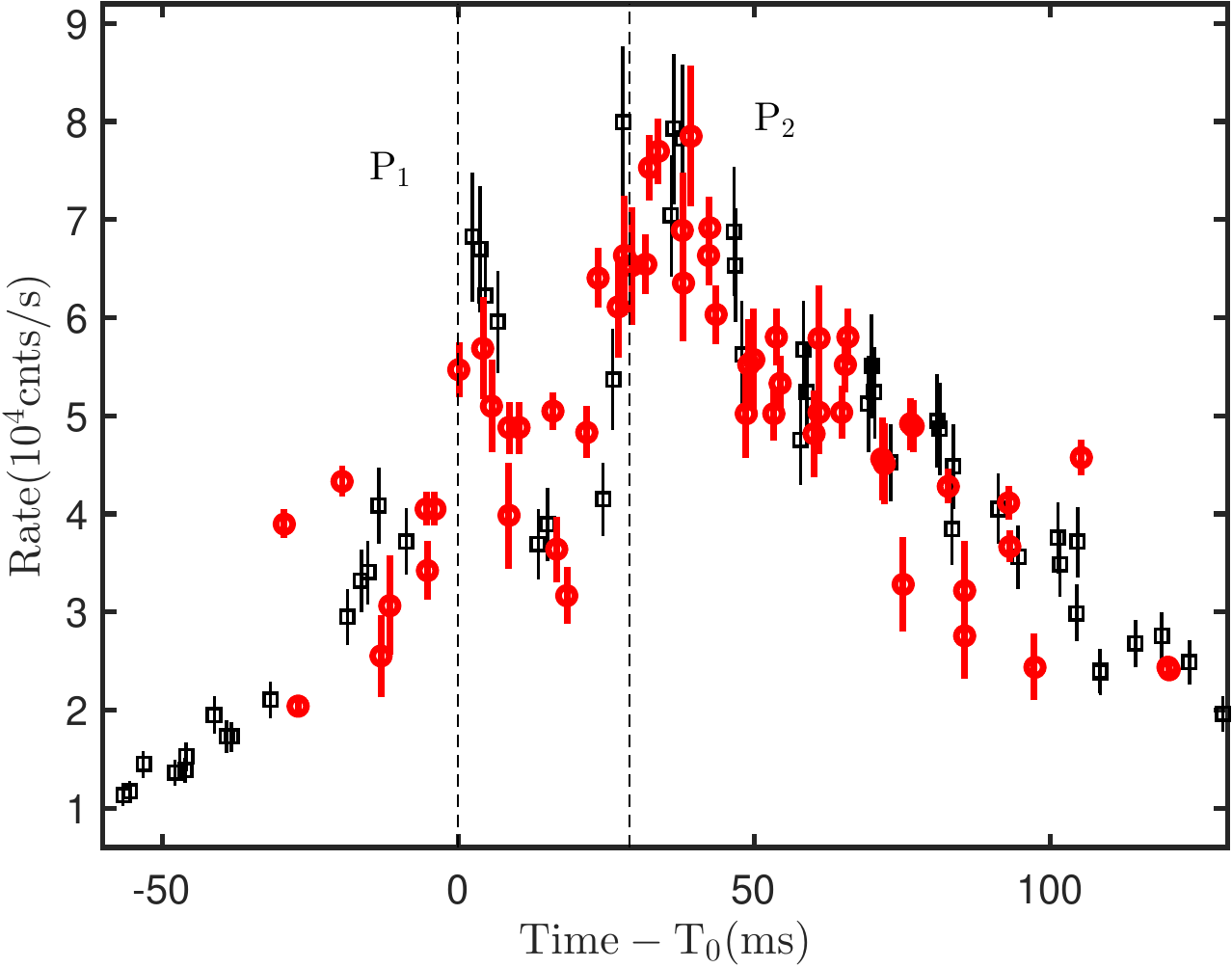}
      }
  \caption{The recovered lightcurve of HE in 15-250\,keV. The black squares and red circles represent the normal and recovered data. The two vertical dashed lines in all panels are the two peak positions of FRB 200428 \citep{2020Natur.587...54C}. $T_{0}$ represents 2020-04-28T14:34:24.4265 (UTC).}
  \label{fig:lc_recover}
\end{figure}

HE consists of eighteen phoswich X-ray detectors. These detectors are divided into three groups. Each group contains six detectors sharing one Physical Data Acquisition Unit (PDAU for short). Each PDAU contains a FIFO with 9-bit width and 2048-byte depth acting as data buffer. A half-full flag will be set when 1024 bytes data are written into the FIFO. Then the PDAU will read out and transfer these data to the solid storage device of the satellite platform. This operation takes about 7\,ms. When HE observes a burst with extremely high flux, the FIFO will rapidly be filled up in less than 7\,ms. Then part of the data can not be recorded and will be lost. It results in some gaps in the original lightcurve of HE. This is the so-called saturation effect in HE. As mentioned by \cite{2021NatAs...5..378L}, the XRB corresponding to FRB 200428A saturated HE seriously. It would affect the timing properties and needs to be corrected. Fortunately, with a deep understanding of the working mechanism of PDAU, we can accurately estimate the average count rate during the gaps in the lightcurve where the event data are lost. Even though we can not recover the lost data event by event, the recovered count rate data points can fill the gaps in the lightcurve and significantly improve the quality of the lightcurve. The saturation and deadtime corrections for all HE data consists of several major steps: (1) recover the data points with average count rate with raw data (level 1B data for HE); (2) find the time intervals where no raw data is lost; (3) obtain the deadtime ratio of each detector at different time intervals; (4) screen the data in the time intervals selected in the first step and then calculate the true count rate of each detector; (5) merge these count rates and calculate their errors. As shown in Figure \ref{fig:lc_recover}, the lightcurves obtained from the three groups of detectors are utilized to analyse the timing parameters and the systematic error is set to 0.06 for the light curves, according to their different responses and backgrounds of these HE detectors \citep{2020JHEAp..27...64L,2020JHEAp..27...14L}. The lightcurves for different energy bands could be generated from similar procedure as described in \cite{2021NatAs...5..378L}. 

After saturation and deadtime corrections, three lightcurves of HE, one in the whole energy band (15--250\,keV) and two narrower energy bands (15--60\,keV and 60--250\,keV), are recovered as shown in Figures \ref{fig:lc_recover}, \ref{fig:lightcurvestructure} and \ref{fig:lightcurvestructure2}, which display more detailed structures than the ones without correction as shown in \cite{2021NatAs...5..378L}. The two hard X-ray bursts associated with FRB 200428 are more significant in the HE lightcurves than that in the previous results. 

\section{Timing and spectral analysis}
\subsection{Lightcurve fitting}
\begin{figure}[!htb]
  \centerline{
      \includegraphics[width=0.4\columnwidth]{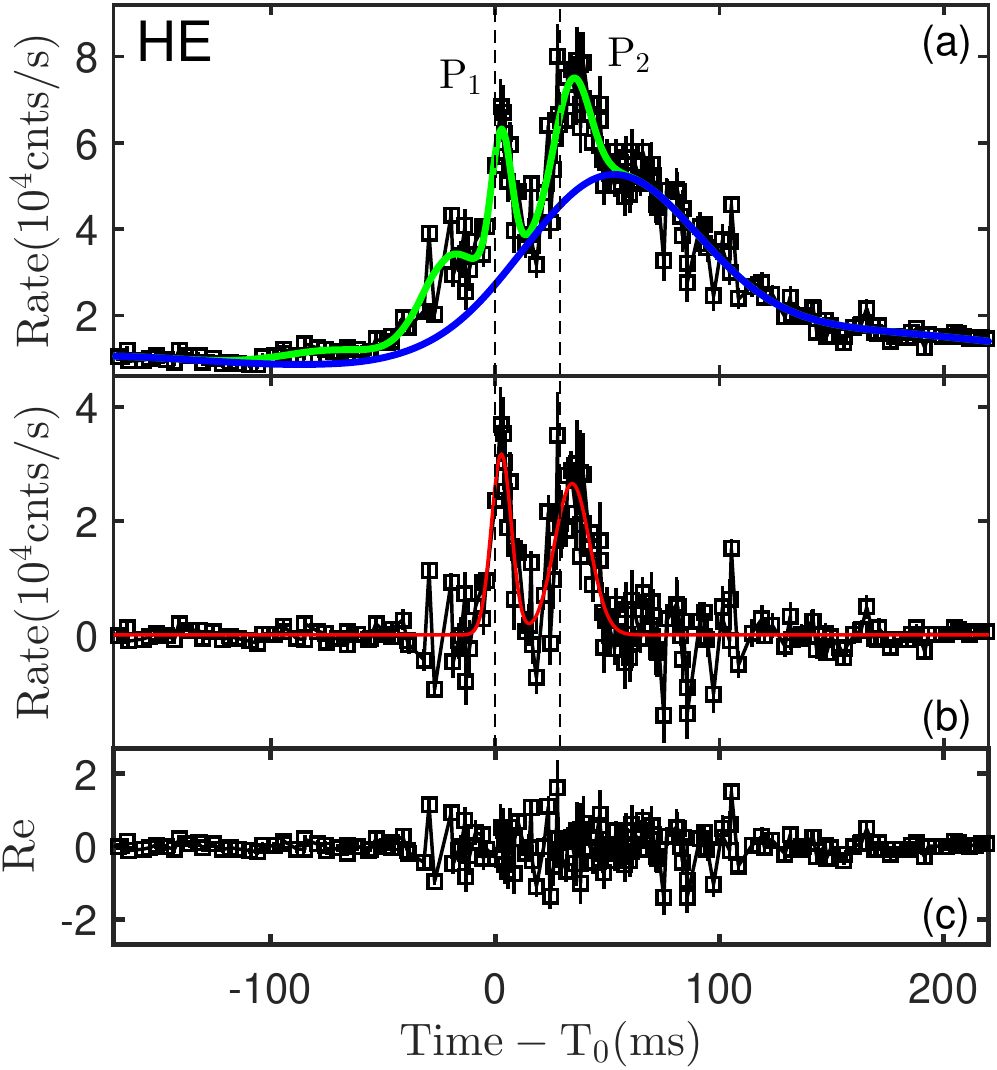}
      \includegraphics[width=0.4\columnwidth]{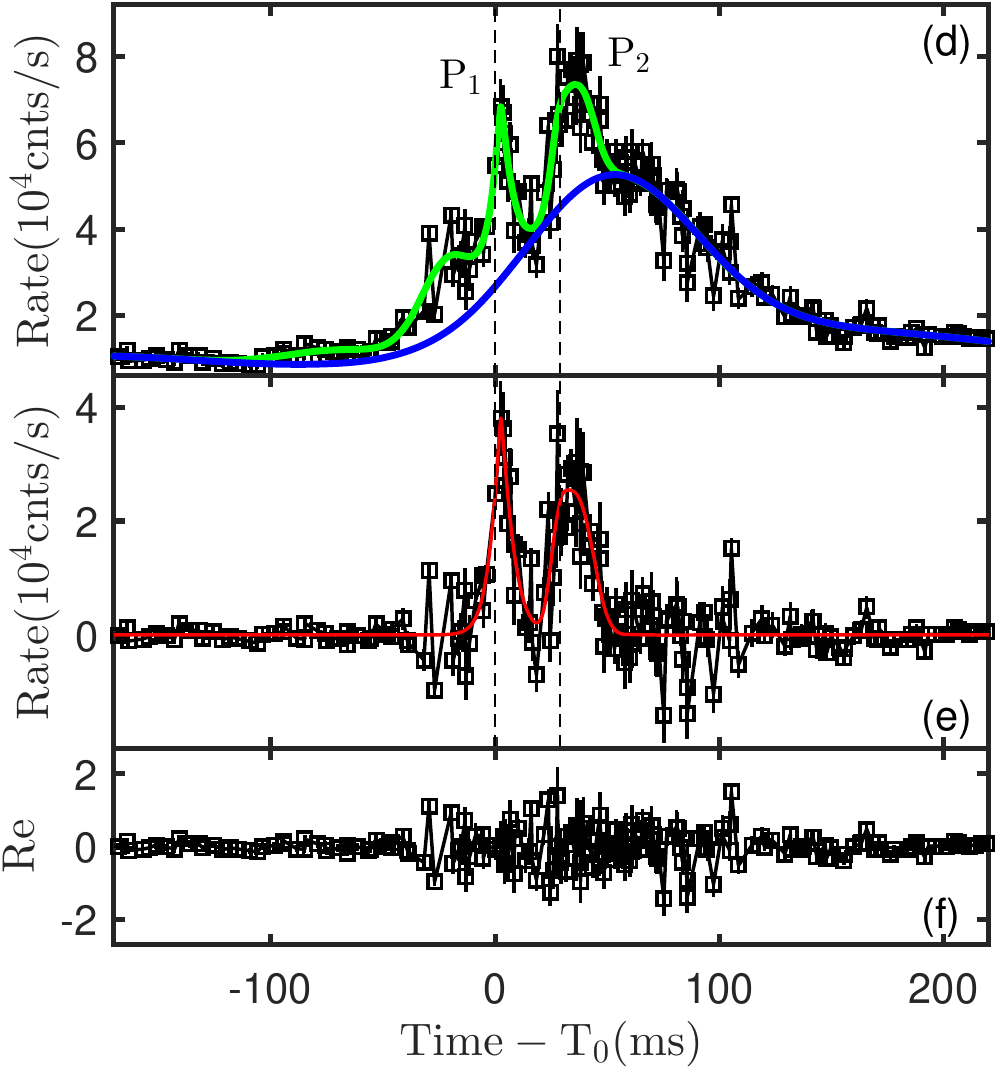}
      }
  \caption{The recovered lightcurves and fitting curvers of HE in energy band 15-250\,keV. Panel (a): The lightcurve of HE after saturation corrections. The green line represents the fitting curve with equation (\ref{fittingFun}) while the blue line only includes the three broad bump-like components. Panel (b): The detrended lightcurve. The red line represents the fitted curve for $\rm{P_{1}}$ and $\rm{P_{2}}$. The two vertical dashed lines in all panels are the two peak positions of FRB 200428 \citep{2020Natur.587...54C}. Panel (c): Residuals between data and the fitting model. Panels (d), (e) and (f) show the fitting results with FRED function for  $\rm{P_{1}}$ and $\rm{P_{2}}$. $T_{0}$ has the identical meaning as in Figure \ref{fig:lc_recover}.}
  \label{fig:lightcurvestructure}
\end{figure}

\begin{figure}[!htb]
  \centerline{
      \includegraphics[width=0.4\columnwidth]{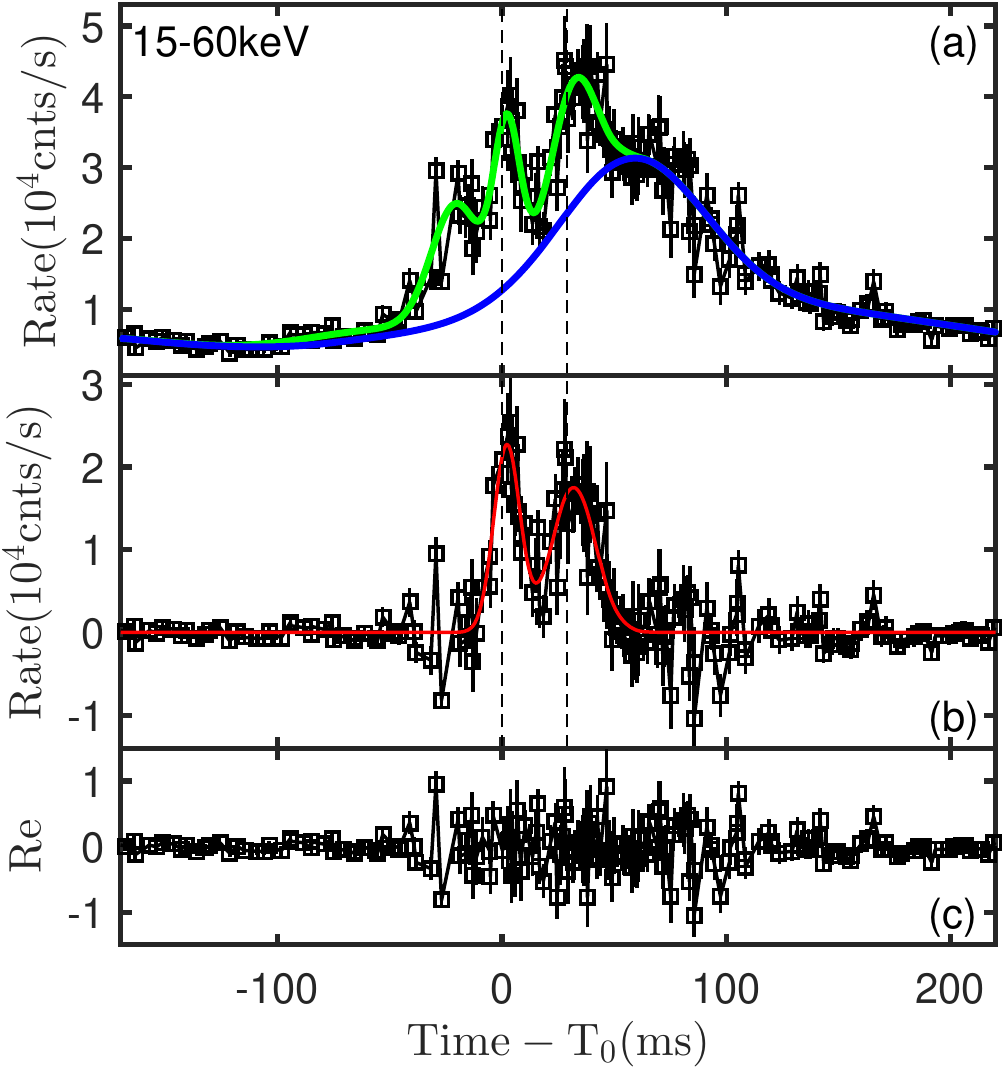}
      \includegraphics[width=0.4\columnwidth]{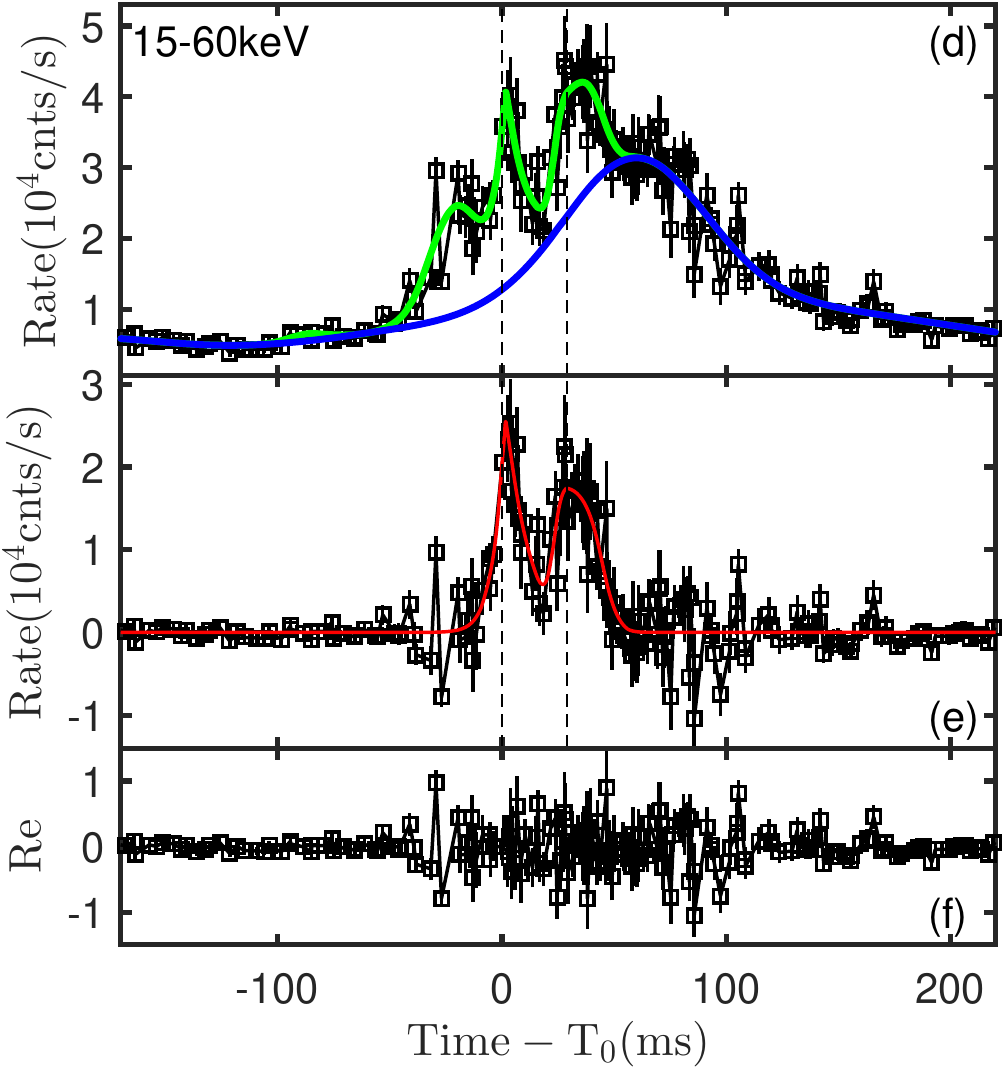}
      }
  \caption{The recovered lightcurves and fitting curvers of HE in energy band 15-60\,keV. Panels (a)-(f) have similar meanings with Figure \ref{fig:lightcurvestructure} but for energy band 15-60\,keV.}
  \label{fig:lightcurvestructure2}
\end{figure}

We utilize several Gaussian functions to fit the three lightcurves of HE to calculate their timing properties as proposed in \cite{2021NatAs...5..378L}. In order to describe the timing properties of FRB 200428 and the XRB conveniently, all arrival times are set relative to $T_{0}=$ 2020-04-28T14:34:24.4265 (UTC, geocentric), corresponding to the first peak of FRB 200428 \citep{2020Natur.587...54C}. As shown in Figures \ref{fig:lightcurvestructure} and \ref{fig:lightcurvestructure2}, the structures of HE lightcurve in energy band 60--250\,keV are less than that in energy bands 15--60\,keV and 15--250\,keV. Briefly, we take the HE lightcurve in energy band 15--250\,keV as an example to describe the fitting process.

As plotted in Figure \ref{fig:lightcurvestructure}, the HE lightcurve in energy band 15--250\,keV consists of three broad bump-like components and two narrow peaks, corresponding to the two radio bursts of FRB 200428. Except for these five components, there are two other weaker peaks (with less significance) before the two narrow peaks. For convenience, we name the two FRB peaks and the corresponding X-ray peaks as $\rm{P_{1}}$ and $\rm{P_{2}}$. The two weaker peaks are named as $\rm{P_{3}}$ and $\rm{P_{4}}$ while the three broad bump-like components are named as $\rm{P_{5}}$ to $\rm{P_{7}}$ as listed in Table \ref{table:clfit2}. However, we only focus on the timing properties of $\rm{P_{1}}$ and $\rm{P_{2}}$ and the other five components are not discussed further in this work. 

To describe these components, we fit the HE lightcurve with seven Gaussian functions and a constant term, which are used to describe $\rm{P_{1}}$ to $\rm{P_{7}}$,
\begin{equation}
    L_{\rm 7}=N_{\rm P_{1}}G(t,t^{\rm{{HE}}}_{\rm P_{1}},w_{\rm P_{1}}) + N_{\rm P_{2}}G(t,t^{\rm{{HE}}}_{\rm P_{2}},w_{\rm P_{2}})+L_{\rm 5}+B_{\rm 1},
    \label{fittingFun}
\end{equation}
where $G(t,t^{\rm{HE}}_{\rm P},w_{\rm P})=\exp{(-(t-t^{\rm HE}_{\rm P})^2/2w^2_{\rm P})}$. $N_{\rm P_{1}}$, $N_{\rm P_{2}}$, $t^{\rm{{HE}}}_{\rm P_{1}}$, $t^{\rm{HE}}_{\rm P_{2}}$, $w_{\rm P_{1}}$ and $w_{\rm P_{2}}$ are the coefficients, the arrival times and Gaussian widths of $\rm{P_{1}}$ and $\rm{P_{2}}$ detected by HE, respectively. $L_{\rm 5}=\sum_{i=3}^{7}{N_{\rm P_{i}}G(t,t^{\rm{HE}}_{\rm P_{i}},w_{\rm P_{i}})}$ is the sum of the five Gaussian functions for describing the two small structures and three broad bump-like components of the lightcurve located at around $T_0-0.084\,{\rm s}$, $T_0+0.011\,{\rm s}$, $T_0-0.17\,{\rm s}$, $T_0+0.05\,{\rm s}$ and $T_0+0.21\,{\rm s}$, respectively. The last term $B_{\rm{l}}$ is the background of the lightcurve.

As shown with the green line in Figure \ref{fig:lightcurvestructure}, the HE lightcurve can be well fitted by equation\,\eqref{fittingFun}. The parameters of the best fitting are listed in Table \ref{table:clfit2} with a reduced $\chi^2$ of 1.45 for 290 degrees of freedom. 

The fitting process of the HE lightcurve in energy band 15-60\,keV is the same as described above. However, the two components $\rm{P_{3}}$ and $\rm{P_{6}}$ are not present when fitting the HE lightcurve in energy band 60--250\,keV as shown in Figure \ref{fig:lightcurvestructure2}(b). The fitting parameters and goodness for these two lightcurves are also listed in Table \ref{table:clfit2}.

The burst shape of FRB 200428 is typically asymmetric with a fast rise and a slow decay. Here, we also utilize a common form called the fast-rising exponential decay (FRED, \cite{1996ApJ...459..393N}) to replace the Gaussian functions for $\rm{P_{1}}$ and $\rm{P_{2}}$ as follows 
\begin{equation}
    I(t)= \begin{cases}N \exp \left[-\left(\frac{\left|t-t_{\max }\right|}{\alpha}\right)^{\gamma}\right] , & t<t_{\max } \\ N \exp \left[-\left(\frac{\left|t-t_{\max }\right|}{\beta}\right)^{\gamma}\right], & t>t_{\max }\end{cases}
    \label{eq-fred}
\end{equation}, where $N$ is the normalization parameter, $t_{\max}$ is the peak time, $\alpha$ and $\beta$ are the rise and decay time constants, $\gamma$ is the sharpness of the pulse for each time bin. The fitting parameters and goodness for these two lightcurves are also listed in Table \ref{table:clfit3}. Limited by the data quality, only the lightcurves of {\it Insight}-HXMT/HE can be fitted with FRED functions.

\subsection{Calculation of timing parameters}
To quantify the timing properties of FRB 200428 and the XRB, the arrival times of all peaks, the time delays between radio peaks and X-ray peaks, peak separation for radio peaks or X-ray peaks are calculated in this work. The arrival times have been obtained by fitting the HE lightcurve, as listed in Tables \ref{table:clfit2} and \ref{table:clfit}. We also utilize the results from CHIME, ME and INTEGRAL and Konus-{\sl Wind}, which are obtained from the literature \citep{2020Natur.587...54C,2021NatAs...5..378L,2020ApJ...898L..29M,2021NatAs...5..372R}. For clarity, we denote $t_{\rm{P_{1}}}^{\rm{\rm{*}}}$ and $t_{\rm{P_{2}}}^{\rm{\rm{*}}}$ as the arrival time of the two peaks, where $\rm{*}=$ FRB, HE, ME, KW, IN or X to represent FRB 200428, HE, ME, Konus-\textit{Wind}, INTEGRAL/IBIS or XRB. Here X represents the weighted mean value for the X-ray results obtained from different X-ray telescopes. After that, the time delays between radio and X-ray peaks are calculated as $\tau_{\rm P_{1}} = t_{\rm P_{1}}^{\rm X}-t_{\rm P_{1}}^{\rm FRB}$ and $\tau_{\rm P_{2}} = t_{\rm P_{2}}^{\rm X}-t_{\rm P_{2}}^{\rm FRB}$, respectively. At the same time, the peak separations for radio or X-ray peaks are calculated with ${S}^{\rm *}=t_{\rm P_{2}}^{\rm *}-t_{\rm P_{1}}^{\rm *}$, where $\rm{*}$ represents the same meanings. Finally, the difference between radio peak separation ${S}^{\rm FRB}$ and X-ray peak separation ${S}^{\rm X}$ is obtained as $\Delta{S} = {S}^{\rm X}-{S}^{\rm FRB}$. The pulse width is defined as the full width at half maximum of the peak. In the end, the values of these parameters are summarised and listed in Table \ref{table:clfit}.

\subsection{Spectra of the two narrow peaks}

\begin{figure}[!htb]
  \centerline{
      \includegraphics[width=0.6\columnwidth]{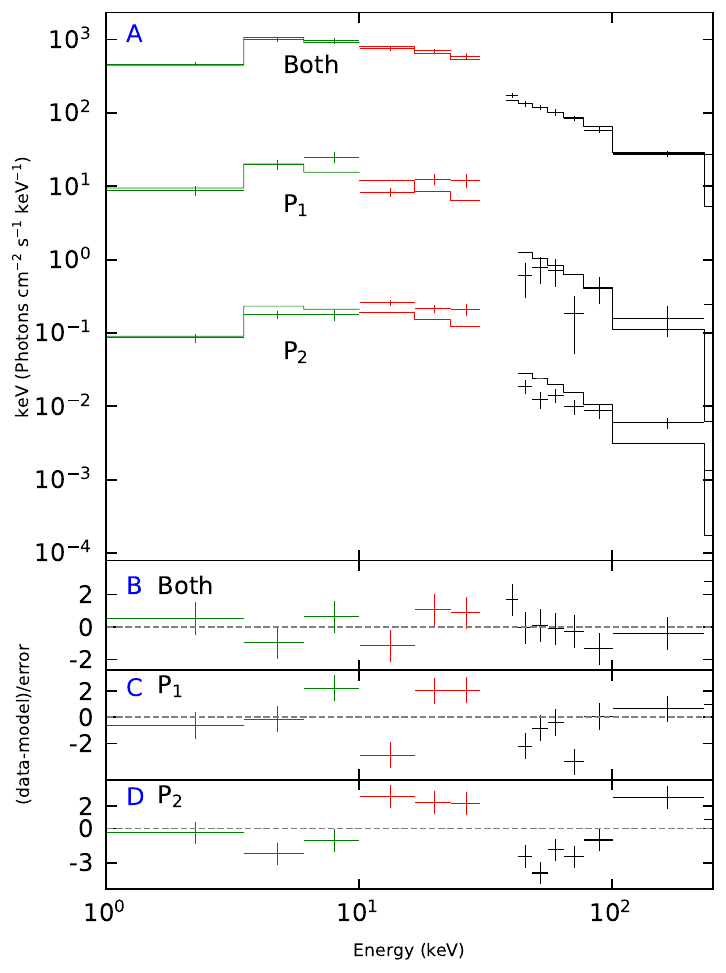}
      }
  \caption{The spectra observed with {\sl Insight}-HXMT covering the 1–250\,keV energy band, with factors 100, 1 and 0.01 for clarity. The black, red and green points represent the data of HE, ME and LE, respectively. The real line is the fitting result of cutoff power-law model.}
  \label{fig:peak_spec}
\end{figure}

In order to obtain the spectra of the $\rm{P_{1}}$ and $\rm{P_{2}}$ individually and combined, we subtract the  background using the background algorithm implemented within the RMFIT and GRM data tool \footnote{\url{https://fermi.gsfc.nasa.gov/ssc/data/analysis/software/}}. Taking HE as an example, the estimation procedure of the background of HE is described as follows: (1) extract the lightcurve for the specific channel; (2) exclude the the time interval of $\rm{P_{1}}$ and $\rm{P_{2}}$; (3) fit the rest lightcurve with Gaussian functions as background; (4) calculate the background for the selected channel. The net spectra of LE and ME are extracted similarly. Finally, the net spectra of $\rm{P_{1}}$, $\rm{P_{2}}$ and both of them are fitted with {\sl cons*TBabs*cutoffpl} and \textit{pgstat} statistics in XSPEC as \cite{2021NatAs...5..378L}. In the fitting process, constant factors are added to implement the different saturation and deadtime effects in different detectors \citep{2021NatAs...5..378L}. Because of the low statistics, we fix the constant factors of $\rm{P_{1}}$ and $\rm{P_{2}}$ to the values derived from the joint fitting of both peaks. The spectral results can be found in Figure \ref{fig:peak_spec} and the best fitting parameters are listed in Table
\ref{table:peak_spec_pars}.

\section{Results and discussions}

\begin{figure}[!htb]
  \centerline{
      \includegraphics[width=0.5\columnwidth]{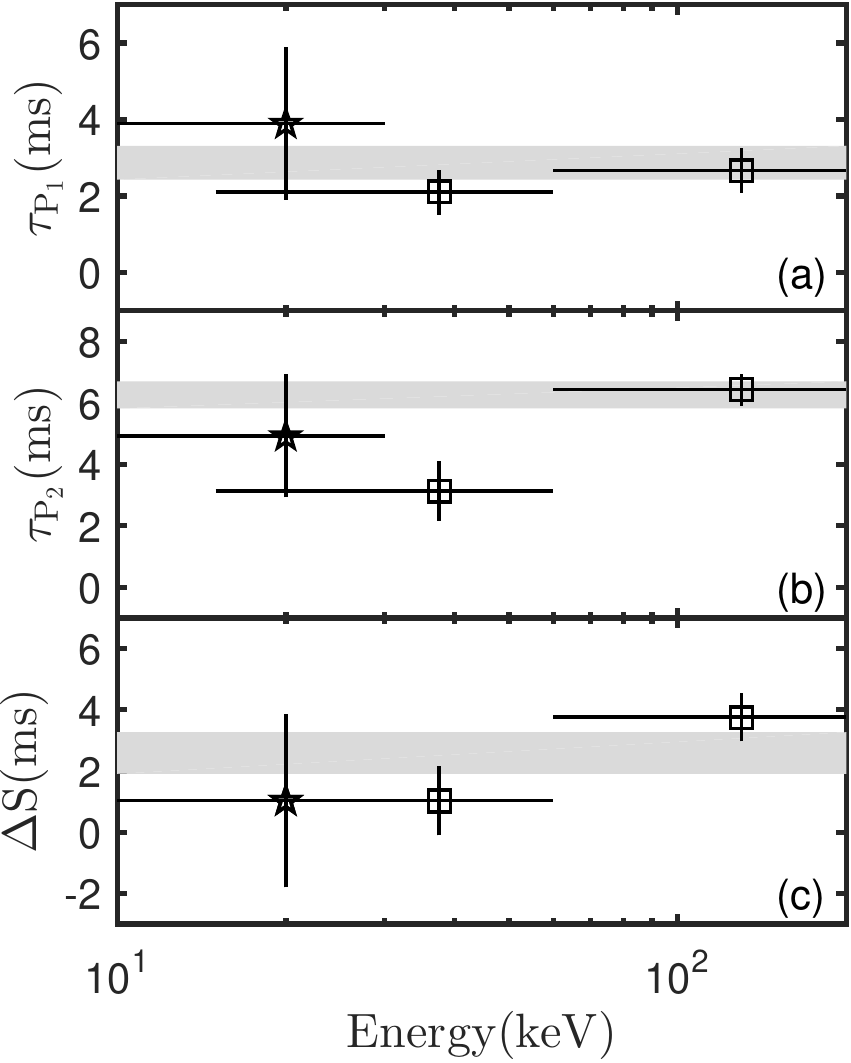}
      }
  \caption{The timing parameters of XRB as function of energy. Panels (a), (b) and (c) show : $\tau_{\rm P_{1}}$, $\tau_{\rm P_{2}}$ and $\Delta{S}$ measured from different instruments, respectively. The pentagrams and squares represent the parameters obtained from \textit{Insight}-HXMT/ME and \textit{Insight}-HXMT/HE, respectively. The grey bands represent the mean values (with 1$\sigma$ uncertainty) of the timing parameters by fitting with multi-Gaussian functions as listed in Table \ref{table:clfit}.}
  \label{fig:lightcurvestructure4}
\end{figure}

The timing parameters measured from different instruments or in different energy bands are plotted in Figure \ref{fig:lightcurvestructure4}. The timing parameters do not show significant evolution as function of energy according to the energy coverage and uncertainties. Considering these factors, the timing parameters could be combined to reduce the statistical errors.

As listed in Table \ref{table:clfit}, the values of the arrival time $t_{\rm P_{1}}$ measured from HE, ME, Konus-{\sl Wind} and INTEGRAL are $2.77\pm0.45$\,ms, $3.9\pm2.0$\,ms, $2.6\pm3.0$\,ms, and $7.5\pm3.0$\,ms, respectively \citep{2021NatAs...5..372R,2020ApJ...898L..29M}. By combining the four arrival times with a weighted average, we obtain $t_{\rm{P_{1}}}^{\rm{X}}=2.92\pm0.43$\,ms. Similarly, we obtain $t_{\rm{P_{2}}}^{\rm{X}}=35.17\pm0.43$\,ms from the combined X-ray observations as listed in Table \ref{table:clfit}. According to the arrival times of radio emission \citep{2020Natur.587...54C}, we can obtain the time delays of the two X-ray peaks as $\tau_{\rm P_{1}}=2.92\pm0.43$\,ms and $\tau_{\rm P_{2}}=6.20\pm0.43$\,ms, respectively. It is the first time that the delay of the X-ray emission of $\rm{P_{1}}$ relative to the first radio burst is obtained with a high confidence level of $6.8\,\sigma$. We also confirm that the X-ray emission of $\rm{P_{2}}$ delays relative to the second radio peak at an even higher confidence level than the previous results, which is longer than that of $\rm{P_{1}}$ \citep{2021NatAs...5..378L,2021NatAs...5..372R,2020ApJ...898L..29M}. 

The separation of the two X-ray peaks ${S}^{\rm{X}}$ is $31.81\pm0.65$\,ms, which is longer than the radio peak separation of ${S}^{\rm{FRB}}=28.97\pm0.02$\,ms detected by CHIME/FRB \citep{2020Natur.587...54C}. Hence, the difference $\Delta{S}$ between the radio peak separation and the X-ray peak separation is $2.84\pm0.65$\,ms, which is above zero at 4.3\,$\sigma$ significance level and is also more significant than that reported in previous results \citep{2021NatAs...5..378L,2021NatAs...5..372R,2020ApJ...898L..29M}. We conclude that there is a difference in the observed peak separations between the two bursts of the FRB and its X-ray counterpart.

\begin{figure}[!htb]
  \centerline{
      \includegraphics[width=0.4\columnwidth]{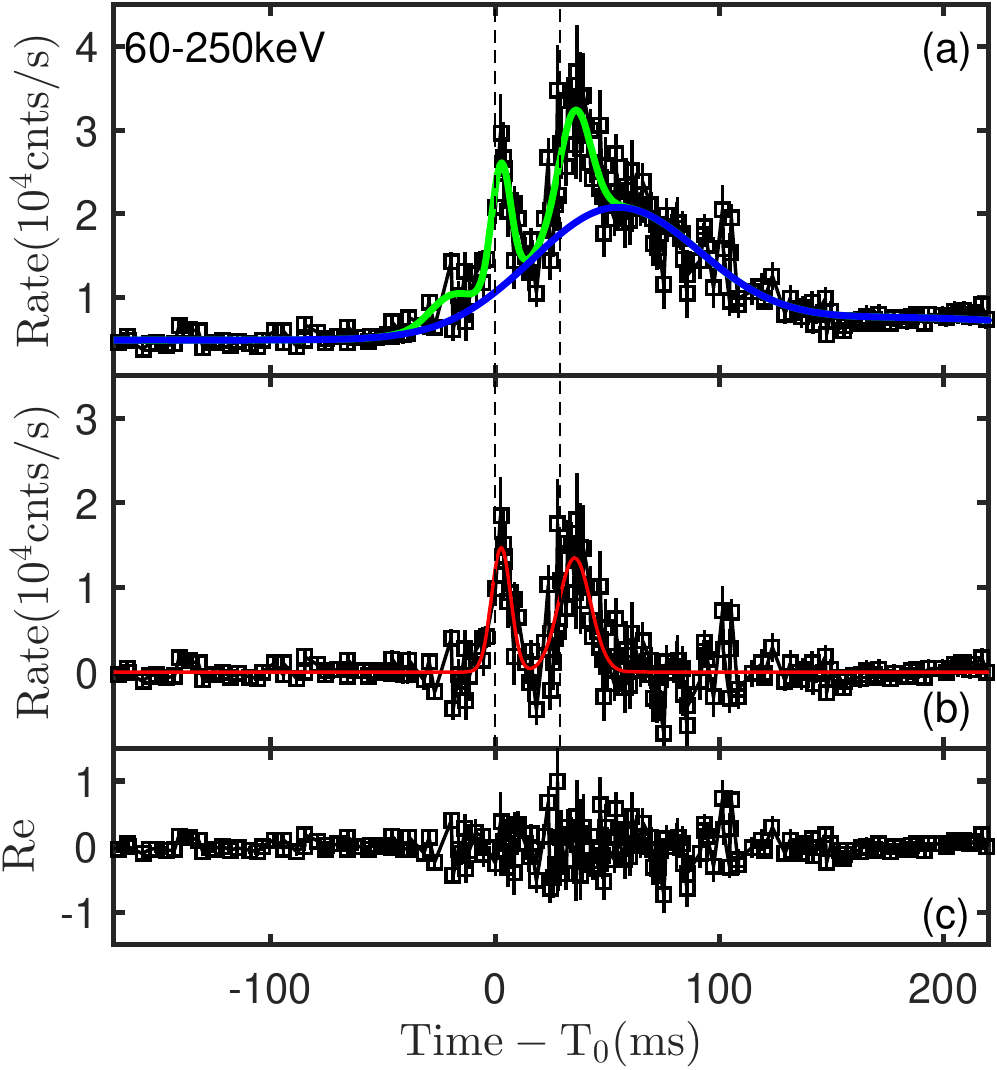}
      \includegraphics[width=0.4\columnwidth]{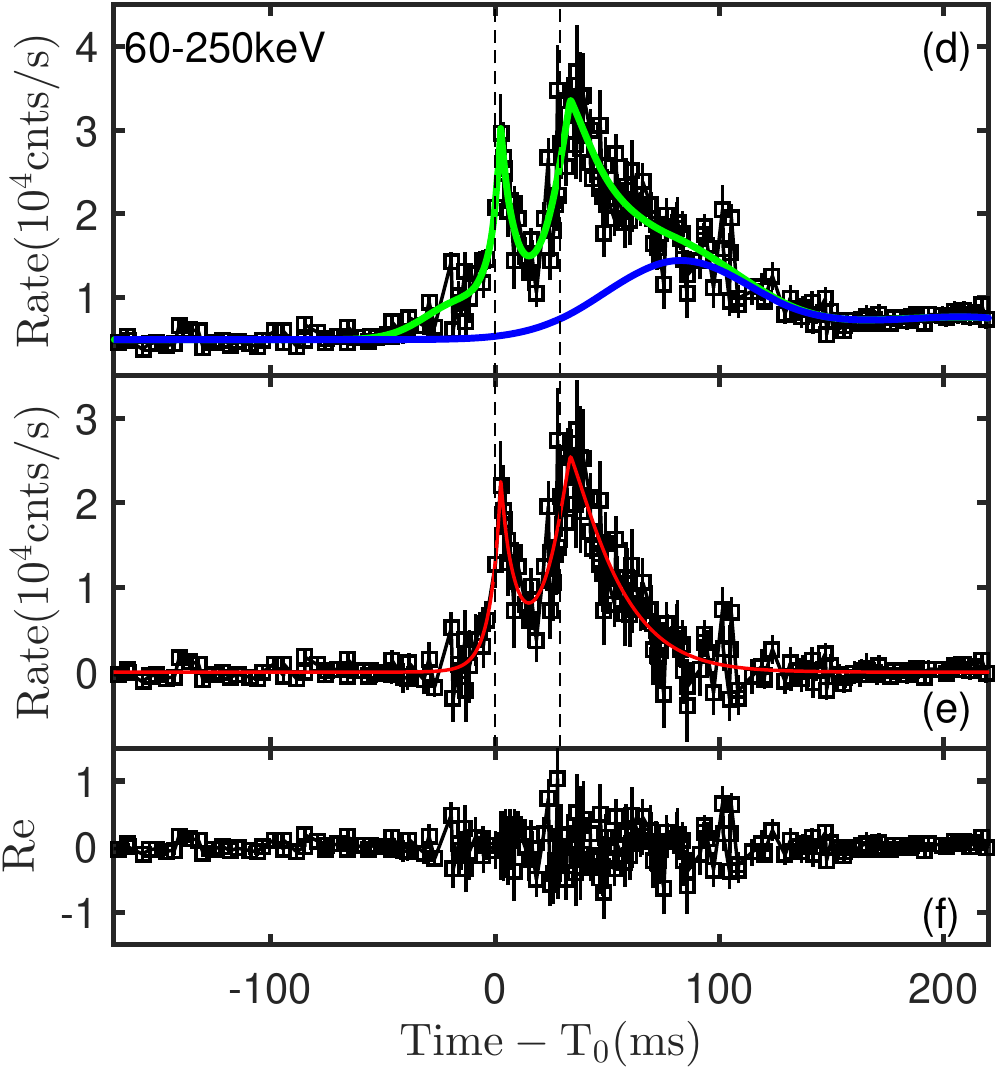}
      }
  \caption{The recovered lightcurves and fitting curves of HE in energy band 60-250\,keV. Panels (a)-(f) have similar meanings with Figure \ref{fig:lightcurvestructure} but for 60-250\,keV.}
  \label{fig:lightcurvestructure3}
\end{figure}

As shown in the right panels of Figures \ref{fig:lightcurvestructure}, \ref{fig:lightcurvestructure2} and \ref{fig:lightcurvestructure3}, 
$\rm{P_{1}}$ and $\rm{P_{2}}$ could be also well fitted by a FRED function for the three lightcurves with energy bands 15-250\,keV, 15-60\,keV and 60-250\,keV. The timing parameters do not evolve with energy from the FRED fitting results and similar with the results fitting with multi-Gaussian functions as listed in Table \ref{table:clfit3}. Considering these factors, the timing parameters are only discussed for the whole energy band of {\sl Insight}-HXMT/HE. With the timing parameters of FRED function, the X-ray peak of $\rm{P_{1}}$ is definitely delayed by about $\tau_{\rm P_{1}}=2.57\pm0.52$\,ms while $\rm{P_{2}}$ shows longer delay of $\tau_{\rm P_{2}}=3.58\pm1.40$\,ms. The separation between the two X-ray peaks is ${S}^{\rm{X}}=30.0\pm1.5$\,ms, which is roughly consistent with ${S}^{\rm{FRB}}$ as this value is
somewhat model (Gaussian or FRED) dependent. Interestingly, in the 60-250\,keV band, one can also get ${S}^{\rm{X}}=31.1\pm0.9$\,ms from the FRED fitting result. Within the margin of error, this value agrees with that of ${S}^{\rm{X}}=32.7\pm0.7$\,ms obtained from the multi-Gaussian fitting result.

\begin{figure}[!htb]
  \centerline{
      \includegraphics[width=0.6\columnwidth]{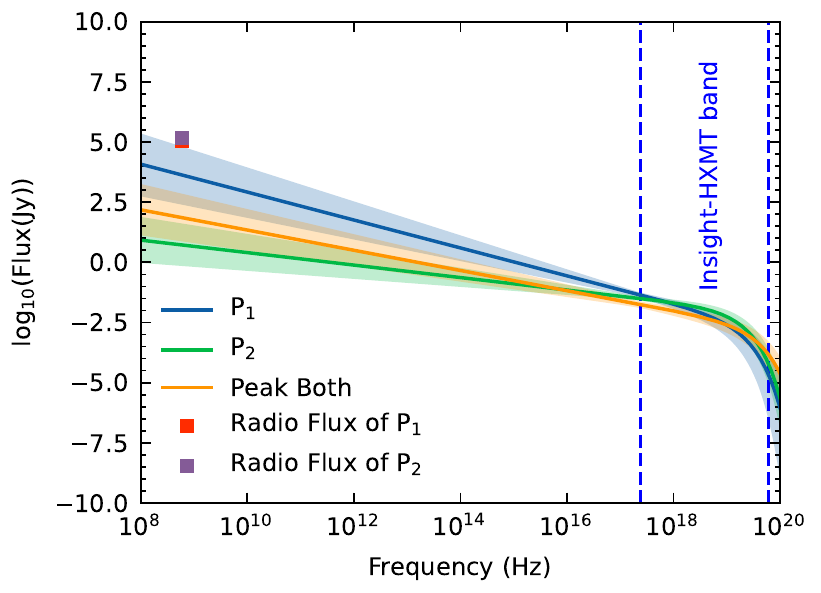}
      }
  \caption{The comparison between the radio flux reported in \citet{2020Natur.587...54C}
   and extrapolations from the X-ray spectra to the radio frequency range,
   where the shadow regions are the 1\,$\sigma$ error bands with the parameters of the cutoff power-law described in Table \ref{table:peak_spec_pars}. The region between two vertical blue dashed lines represents $Insight$-HXMT energy band.}
  \label{fig:peak_spec_comp}
\end{figure}

We also obtained the detailed spectral information of $\rm{P_{1}}$ and $\rm{P_{2}}$ as shown in Figure \ref{fig:peak_spec} and Table \ref{table:peak_spec_pars}. From the fitting results, the photon index $1.4\pm0.1$ of the combined spectrum of $\rm{P_{1}}$ and $\rm{P_{2}}$ is softer than the previous result of $0.9\pm0.1$ of the spectrum including both the two bursts and the broader components of the burst in the same time interval when the two bursts occurred \citep{2021NatAs...5..378L}. The cutoff energy is consistent with the previous result within $1.6\,\sigma$ errors. In addition, the photon index and cutoff energy of $\rm{P_{1}}$ are consistent with $\rm{P_{2}}$ as listed in Table \ref{table:peak_spec_pars}. Finally, the radio fluxes are higher than that extrapolated from X-ray spectral parameters as shown in Figure \ref{fig:peak_spec_comp} \citep{2021NatAs...5..378L,2020Natur.587...54C}.

Many recent theoretical models attribute FRBs and XRBs to various sorts of instablities in the magnetar's magnetosphere \citep{2020ApJ...900L..26W,2019ApJ...886..110M,2020ApJ...904L...5X,2020ApJ...899L..27M,2021MNRAS.500.2704Y,2020ApJ...900L..21Y,2019ApJ...879....4W,2020MNRAS.494.2385K,2020ApJ...901L..13Y,2020MNRAS.498.1397L,2021ApJ...919...89Y,2020ApJ...897L..40D,2020ApJ...898L..55G}. In these scenarios, the instabilities are triggered by perturbations in a certain critical condition somewhere in the magnetosphere where FRBs and XRBs are emitted. Indeed, these scenarios can be appllied in the most general cases, although there may be differences in the origin between various trigger mechanisms.

In general, the observed time delay between any two burst events ${\rm EVT_{1}}$ and ${\rm EVT_{2}}$ can be given by
\begin{equation}
\Delta\,\!\tau_{\rm{obs}}=\frac{~d_{2}-d_{1}~}{c}+t_{\rm EVT_{2}}-t_{\rm EVT_{1}},
\label{eq:tdelay}
\end{equation}
where $d_{1}$ and $d_{2}$ are the distances of the emitting points to the observer on the Earth, and $t_{\rm EVT_{1}}$ and $t_{\rm EVT_{2}}$ are the moments when the bursts emerge. Here, ${\rm EVT_{1}}$ and ${\rm EVT_{2}}$ can be regarded as the two peaks in the XRB. There may be three possibilities. First, if the two events are caused by individual perturbations successively occurred at the same place or in a very small area, the time interval of the two events dominates the time delay, $\Delta\,\!\tau_{\rm{obs}} = (t_{\rm EVT_{2}}-t_{\rm EVT_{1}})=S^{\rm{X}}$ for the XRB. Similarly, we have $\Delta\,\!\tau_{\rm{obs}}=S^{\rm{FRB}}$ for the FRB. 
Second, the two emission events may be thought of as originated from the same perturbation with two pulses. 
In this case, the time delay between the radio and X-ray peaks is dominated by perturbation propagation if we suppose that the propagation speed is much less than the light speed. We have $\Delta\,\!\tau_{\rm{obs}}=\tau_{\rm{P_{1}}}~{\rm or}~\tau_{\rm{P_{2}}}$. Third, it is also possible that the two burst events corresponding to the two peaks in the XRB/FRB occur independently in two fairly distant places. In this case, the two burst events are totally incidental coincidence in tens of milliseconds.

From the results of timing analysis, we can conclude that the first X-ray peak is delayed by about 3\,ms relative to the first peak of FRB 200428. The separation between the two X-ray peaks  may or may not show significant difference with the separation of the two radio peaks of FRB 200428, depending on which model (Gaussian or FRED) is applied in fitting the lightcurves of the two X-ray peaks. On the other hand, the time-resolved spectral analysis shows that there are no significant differences in the spectral parameters of the two XRB sub-bursts $\rm{P_{1}}$ and $\rm{P_{2}}$, such as the photon index $\alpha\sim1.4$, and cutoff energy $E_{\rm{cut}}\sim60$\,keV, which indicates the same origin for the two sub-bursts. Actually, the spectral parameters of the two sub-bursts in the XRB are also consistent with the trend of the overall spectral evolution during the XRB \citep{2021NatAs...5..378L,2020ApJ...898L..29M,2021NatAs...5..372R}.

Combining the results of timing analysis and spectral analysis, it is very likely that the two sub-bursts $\rm{P_{1}}$ and $\rm{P_{2}}$ of FRB 200428 and its associated XRB should originate from the same circumburst environment, or be affected by the same environmental factors. We thus suggest that the FRB and XRB originated from the same perturbation with two sub-bursts, emitting radio and X-ray at different places satisfying the respective critical condition of instability in the propagation of the perturbation, as the second case discussed above.

\section{Summary}

In this work, we present a more detailed timing analysis to the lightcurves of \textit{Insight}-HXMT/HE observation of the XRB associated with FRB 200428 after the saturation corrections. The arrival times of the two X-ray peaks by fitting the lightcurve with multi-Gaussian functions are $2.77\pm0.45$\,ms and $34.30\pm0.56$\,ms, which are much more accurate than the previous results. Together with the results of \textit{Insight}-HXMT/ME, INTEGRAL and Konus-{\sl Wind}, the weighted mean time delays of the two X-ray peaks corresponding to radio peaks are $2.92\pm0.43$\,ms and $6.21\pm0.43$\,ms, respectively. Then, we also obtained the peak separation between the two X-ray peaks of $31.81\pm0.65$\,ms, which is longer than that of the two radio peaks of FRB 200428 with $2.84\pm0.65$\,ms. At the same time, the lightcurve of \textit{Insight}-HXMT/HE is fitted with an FRED function since the peak shape is asymmetric. According to FRED fitting results, the time delays of two X-ray peaks are $2.57\pm0.52$\,ms and $3.58\pm1.40$\,ms while the peak separation is equal to or greater than the separation of the two radio peaks, which is model dependent. In addition, these parameters do not show significant evolution with energy considering their errors. Finally, the spectrum of the two X-ray peaks are similar, which could be fitted by cutoff power-law with photon index $\sim{1.4}$ and cutoff energy $\sim{60}$\,keV. These results could supply more constraints on the emission models of FRBs.

\section*{Acknowledgments}
This work is supported by the National Key R\&D Program of China (2021YFA0718500) from the Minister of Science and Technology of China (MOST). The authors thank supports from the National Natural Science Foundation of China under Grants 12173103, U2038101, U1938103 and 11733009. This work is also supported by International Partnership Program of Chinese Academy of Sciences (Grant No.113111KYSB20190020).



\clearpage

\clearpage
\begin{table}
\footnotesize
\caption{Fitting parameters of HE lightcurves with multi-Gaussian functions}
\scriptsize{}
\label{table:clfit2}
\medskip
\begin{center}
\begin{tabular}{l l c c c}
\hline\hline
Peak No. & Parameters & All & 15--60\,keV & 60--250\,keV \\
\hline
&$N_{\rm P_{1}}$ ($10^{3}$\,cnts/s) &$31.8\pm3.1$ & $22.6\pm3.3$ & $14.8\pm1.3$\\
$\rm{P_{1}}$ &$t^{\rm{HE}}_{\rm P_{1}}$ (ms)& $2.77\pm0.45$ & $2.14\pm0.6$ & $2.67\pm0.50$\\
&$w_{\rm P_{1}}$ (ms) &$4.44\pm0.49$ &$5.84\pm0.67$ &$4.26\pm0.50$\\
\hline
&$N_{\rm P_{2}}$ ($10^{3}$\,cnts/s) &$26.6\pm2.2$ & $17.5\pm4.2$ & $13.5\pm1.0$\\
$\rm{P_{2}}$ &$t^{\rm{HE}}_{\rm P_{2}}$ (ms)& $34.3\pm0.56$ & $31.9\pm0.94$ & $35.4\pm0.53$\\
&$w_{\rm P_{2}}$ (ms) &$7.82\pm0.79$ &$9.79\pm1.7$ &$7.12\pm0.67$\\
\hline
&$N_{\rm P_{3}}$ ($10^{3}$\,cnts/s) &$3.2\pm3.6$ & $1.22\pm7.6$ & $-$\\
$\rm{P_{3}}$ &$t^{\rm{HE}}_{\rm P_{3}}$ (ms)& $-73.7\pm16.0$ & $-66\pm57$ & $-$\\
&$w_{\rm P_{3}}$ (ms) &$21.9\pm22$ &$28\pm96$ & $-$\\
\hline
&$N_{\rm P_{4}}$ ($10^{3}$\,cnts/s) &$16.7\pm3.0$ & $15.9\pm3.1$ & $3.0\pm1.3$\\
$\rm{P_{4}}$ &$t^{\rm{HE}}_{\rm P_{4}}$ (ms)& $-21.1\pm2$ & $-21\pm1.5$ & $-20.4\pm3.9$\\
&$w_{\rm P_{4}}$ (ms) &$12.2\pm2.9$ &$10.7\pm2.2$ &$9.25\pm4.8$\\
\hline
&$N_{\rm P_{5}}$ ($10^{3}$\,cnts/s) &$38.7\pm7.3$ & $20.8\pm6.3$ & $15\pm1.3$\\
$\rm{P_{5}}$ &$t^{\rm{HE}}_{\rm P_{5}}$ (ms)& $50.9\pm3.0$ & $58.1\pm8.1$ & $53.5\pm2.7$\\
&$w_{\rm P_{5}}$ (ms) &$38.5\pm4.2$ &$32.3\pm7.2$ &$37.6\pm3.2$\\
\hline
&$N_{\rm P_{6}}$ ($10^{3}$\,cnts/s) &$3.93\pm1.1$ & $3.64\pm2.5$ & $-$\\
$\rm{P_{6}}$ &$t^{\rm{HE}}_{\rm P_{6}}$ (ms)& $-183\pm25$ & $-189\pm29$ & $-$\\
&$w_{\rm P_{6}}$ (ms) &$56.1\pm25.0$ &$44.4\pm22.0$ & $-$\\
\hline
&$N_{\rm P_{7}}$ ($10^{3}$\,cnts/s) &$10.1\pm2.6$ & $9.12\pm3.9$ & $2.66\pm0.41$\\
$\rm{P_{7}}$ &$t^{\rm{HE}}_{\rm P_{7}}$ (ms)& $137\pm68$ & $90.6\pm97.2$ & $180\pm39$\\
&$w_{\rm P_{7}}$ (ms) &$105\pm41$ &$118\pm46$ &$44.4\pm36.0$\\
\hline
Bkg &$B_{\rm{l}}$($10^{3}$\,cnts/s)& $6.59\pm0.55$ & $1.80\pm0.34$ & $4.88\pm0.18$\\
\hline
&$\chi^2({\rm d.o.f})$ &1.45(290)&1.56(290)&1.58(296)\\
\hline
\end{tabular}
\end{center}
\end{table}

\clearpage

\begin{table}
\footnotesize
\caption{Fitting parameters of HE lightcurves with FRED function}
\scriptsize{}
\label{table:clfit3}
\medskip
\begin{center}
\begin{tabular}{l l c c c}
\hline
\hline
Peak No. & Parameters & All & 15--60\,keV & 60--250\,keV \\
\hline
&$N_{\rm P_{1}}$ ($10^{3}$\,cnts/s) &$38.3\pm8.1$ & $25.7\pm5.6$ & $21.7\pm11$\\
$\rm{P_{1}}$ &$t^{\rm{HE}}_{\rm P_{1}}$ (ms)& $2.57\pm0.52$ & $1.58\pm0.68$ & $2.54\pm0.46$\\
&$\alpha_{\rm P_{1}}$ (ms) &$5.15\pm1.6$ &$5.48\pm2.2$ &$4.52\pm3.1$\\
&$\beta_{\rm P_{1}}$ (ms) &$6.06\pm1.7$ &$10.7\pm3.2$ &$6.73\pm3.7$\\
&$\gamma_{\rm P_{1}}$ &$1.19\pm0.4$ &$1.11\pm0.38$ &$0.933\pm0.5$\\
\hline
&$N_{\rm P_{2}}$ ($10^{3}$\,cnts/s) &$25.4\pm3.8$ & $16\pm2.8$ & $25.1\pm11$\\
$\rm{P_{2}}$ &$t^{\rm{HE}}_{\rm P_{2}}$ (ms)& $32.5\pm1.4$ & $30.5\pm1.6$ & $33.6\pm0.78$\\
&$\alpha_{\rm P_{2}}$ (ms) &$8.61\pm1.9$ &$8.39\pm2.2$ &$11.7\pm4.5$\\
&$\beta_{\rm P_{2}}$ (ms) &$13.5\pm1.7$ &$16.2\pm2.1$ &$24\pm20$\\
&$\gamma_{\rm P_{2}}$ &$2.81\pm0.68$ &$2.96\pm0.78$ &$1.18\pm0.35$\\
\hline
&$N_{\rm P_{3}}$ ($10^{3}$\,cnts/s) &$3.15\pm3.9$ & $1\pm1.4$ & $-$\\
$\rm{P_{3}}$ &$t^{\rm{HE}}_{\rm P_{3}}$ (ms)& $-72.3\pm19$ & $-86.4\pm13$ & $-$\\
&$w_{\rm P_{3}}$ (ms) &$23.1\pm25$ &$8.4\pm15$ & $-$\\
\hline
&$N_{\rm P_{4}}$ ($10^{3}$\,cnts/s) &$16.8\pm3.5$ & $15.1\pm1.6$ & $4.2\pm3$\\
$\rm{P_{4}}$ &$t^{\rm{HE}}_{\rm P_{4}}$ (ms)& $-21.5\pm2.3$ & $-21\pm2.1$ & $-15.5\pm26$\\
&$w_{\rm P_{4}}$ (ms) &$11.7\pm3.4$ &$11.2\pm2.3$ &$16.2\pm14$\\
\hline
&$N_{\rm P_{5}}$ ($10^{3}$\,cnts/s) &$38.5\pm7.7$ & $20.2\pm1.6$ & $9.28\pm8.3$\\
$\rm{P_{5}}$ &$t^{\rm{HE}}_{\rm P_{5}}$ (ms)& $51.4\pm4.7$ & $59.1\pm5.4$ & $82\pm29$\\
&$w_{\rm P_{5}}$ (ms) &$38.1\pm5.2$ &$31.2\pm4.4$ &$32.6\pm16$\\
\hline
&$N_{\rm P_{6}}$ ($10^{3}$\,cnts/s) &$3.91\pm1.1$ & $3.44\pm0.72$ & $-$\\
$\rm{P_{6}}$ &$t^{\rm{HE}}_{\rm P_{6}}$ (ms)& $-183\pm26$ & $-192\pm11$ & $-$\\
&$w_{\rm P_{6}}$ (ms) &$55.8\pm25$ &$41.3\pm11$ & $-$\\
\hline
&$N_{\rm P_{7}}$ ($10^{3}$\,cnts/s) &$10.2\pm2.7$ & $9.55\pm0.99$ & $2.68\pm0.45$\\
$\rm{P_{7}}$ &$t^{\rm{HE}}_{\rm P_{7}}$ (ms)& $135\pm71$ & $81.8\pm12$ & $206\pm19$\\
&$w_{\rm P_{7}}$ (ms) &$106\pm42$ &$121\pm12$ &$41.3\pm20$\\
\hline
Bkg &$B_{\rm{l}}$($10^{3}$\,cnts/s)& $6.59\pm0.54$ & $1.79\pm0.32$ & $4.94\pm0.17$\\
\hline
&$\chi^2({\rm d.o.f})$ &1.46(286)&1.54(286)&1.59(292)\\
\hline
\end{tabular}
\end{center}
\end{table}

\clearpage
\begin{table}
\footnotesize
\caption{The timing parameters of FRB 200428 and its X-ray counterpart}
\scriptsize{}
\label{table:clfit}
\medskip
\begin{center}
\begin{tabular}{l | c | c c c c | c | c}
\hline
\hline
Parameters & FRB$^{\rm (a)}$ &  HE$^{\rm (b)}$ &  ME$^{\rm (c)}$  &  KW$^{\rm (d)}$ & IN$^{\rm (e)}$ & X$^{\rm (f)}$ & HE$^{\rm (g)}$ \\
\hline
$t_{\rm{P_{1}}}$ (ms)& $0.00\pm0.02$ & $2.77\pm0.45$ & $3.9\pm2.0$ & $2.6\pm3.0$& $7.5\pm3.0$ & $2.92\pm0.43$ & $2.57\pm0.52$ \\
$\delta_{\rm P_{1}}$ (ms)& $0.585 \pm 0.014$ & $10.4\pm1.1$ &--& --&--&-- & $7.95\pm2.1$\\
$t_{{\rm P_{2}}}$ (ms) & $28.97\pm0.02$& $34.30\pm0.56$ & $33.9\pm2.0$&$37.9\pm1.0$& $35.5\pm1.0$& $35.18\pm0.43$ & $32.5\pm1.4$\\
$\delta_{\rm P{2}}$ (ms)&$0.335 \pm 0.007$ & $18.4\pm1.9$ & --&-- & --& -- & $19.3\pm3.4$ \\
${S}$ (ms) & $28.97\pm0.03$& $31.60\pm0.72$ & $30\pm3$ & $35.3\pm3.2$ &$28.0\pm3.2$&$31.81\pm0.65$ & $30.0\pm1.5$\\
\hline
$\tau_{\rm P_{1}}$ (ms) &-- &$2.77\pm0.45$ & --&-- &--&$2.92\pm0.43$ & $2.57\pm0.52$\\
$\tau_{\rm P_{2}}$ (ms) &-- &$5.33\pm0.56$ & --&-- &--&$6.21\pm0.43$ & $3.58\pm1.40$\\
\hline
$\Delta{S}$ (ms) &-- &$2.63\pm0.72$ & --&-- &--&$2.84\pm0.65$ & $1.0\pm1.5$\\
\hline
\end{tabular}
\\
(a) FRB 200428: \cite{2020Natur.587...54C}.\\
(b) \textit{Insight}-HXMT/HE: parameters fitted by multi-Gaussian functions in this work.\\
(c) \textit{Insight}-HXMT/ME: \cite{2021NatAs...5..378L}\\
(d) INTEGRAL/IBIS: \cite{2020ApJ...898L..29M}.\\
(e) Konus-\textit{Wind}: \cite{2021NatAs...5..372R}.\\
(f) X: the weighted mean values of parameters obtained from several instruments mentioned above.\\
(g) \textit{Insight}-HXMT/HE: parameters fitted by FRED function in this work\\
\end{center}
\end{table}

\clearpage

\begin{table}
\footnotesize
\caption{Fitting parameters of spectra of $\rm{P_{1}}$, $\rm{P_{2}}$ and both of them}
\scriptsize{}
\label{table:peak_spec_pars}
\medskip
\begin{center}
\begin{tabular}{l   c   c   c}
\hline
\hline
Parameter & Both & $\rm{P_{1}}$ & $\rm{P_{2}}$ \\
\hline
Norm & $27.3^{+8.1}_{-6.3}$ & $68.2^{+16.2}_{-13.6}$ & $48.5^{+9.3}_{-8.0}$ \\
PhoIndex & $1.4^{+0.1}_{-0.1}$ & $1.6^{+0.1}_{-0.1}$    & $1.3^{+0.1}_{-0.1}$ \\
HighECut (keV) &$100.3^{+24.8}_{-17.2}$& $56.1^{+16.2}_{-13.6}$ & $51.4^{+11.3}_{-8.2}$ \\
Factor of HE  & $0.4^{+0.04}_{-0.04}$ & 0.4(fixed) & 0.4(fixed) \\
Factor of ME &  1.0(fixed) & 1.0(fixed) & 1.0(fixed) \\
Factor of LE & $0.9^{+0.13}_{-0.12}$ & 0.9(fixed) & 0.9(fixed) \\
$\chi^2/d.o.f.$ & 14.8/9 & 38.1/10 & 57.9/10 \\
\hline
\end{tabular}
\end{center}
\end{table}

\clearpage

\bibliography{msNote}{}
\bibliographystyle{aasjournal}

\end{document}